\documentclass[a4paper, 12pt]{article}

\usepackage{amssymb}
\usepackage{amsmath, epsfig, subfigure,color}
\usepackage{units}

\newcommand{\la}{\label}
\newcommand{\be}{\begin{equation}}
\newcommand{\en}{\end{equation}}

\renewcommand{\vec}[1]{\boldsymbol{#1}}
\newcommand{\ii}{\textrm{i}}
\newcommand{\ee}{\textrm{e}}

\begin{document}

\title{Compression Instabilities of Tissues\\ with Localized Strain Softening}

\author{Michel DESTRADE$^a$, Jose  MERODIO$^b$\\
$^a$School of Mathematics, Statistics, and Applied Mathematics,\\
 National University of Ireland Galway, \\
 University Road, Galway, Ireland. \\[12pt]
 Department of Continuum Mechanics and Structures, \\
E.T.S. Ing. Caminos, Canales y Puertos, \\
Universidad Polit\'ecnica de Madrid, \\Madrid, Spain.}

\date{}

\maketitle

\begin{abstract}

The stress-strain relationship of biological soft tissues affected by Marfan's syndrome is believed to be non-convex.
More specifically, Haughton and Merodio 
recently proposed a strain-energy density leading to localized strain softening, in order to model the unusual mechanical behavior of these isotropic, incompressible tissues.
Here we investigate how this choice of strain energy affects the results of some instabilities studies, such as those concerned with the compression of infinite and semi-infinite solids, slabs, and cylinders, or with the bending of blocks, and draw comparisons with  known results established previously for the case of a classical neo-Hookean solid.
We find that the localized strain softening effect leads to early instability only when instability occurs at severe compression ratios for neo-Hookean solids, as is the case for bulk, surface, and bending instabilities.

\end{abstract}



\newpage


\section{Introduction}


The Marfan pathology is a connective tissue disorder caused by a defect in the fibrillin glicoprotein, a major and essential component of all connective tissues. 
As a result, the arterial waLSS of humans with Marfan syndrome exhibit weaker mechanical properties and behavior compared to healthy arterial waLSS. 
Dilatation of the arteries may further weaken the inner part of the artery wall and may eventually cause tearing and even sudden rupture of the artery. 
It is well documented that because of connective tissue changes, individuals with Marfan syndrome have high risk of aortic dilatation and dissection caused by increased blood pressure and heart rate [Mizuguchi and Matsumoto, 2007].  
Other problems occurring in the aorta of patients with Marfan syndrome include the development of aneurysms, where the artery wall thickens and the artery diameter increases [Watton \emph{et al.}, 2004;  Watton and Hill, 2009] .

Preliminary results on biological soft tissues affected by Marfan syndrome reveal striking differences with healthy tissues, including \emph{(i)} an \emph{isotropic} mechanical behavior and \emph{(ii)} non-monotone uniaxial stress-strain curves with a double curvature. This latter phenomenon can be interpreted as a limited amount of strain-stiffening followed by strain-softening.
It has been called ``\emph{localized strain softening}'' behavior, denoted with the abbreviation LSS. 
In contrast, healthy arterial wall tissues are typically modeled with just strain-stiffening behavior [Holzapfel \emph{et al.}, 2000] and have a highly anisotropic character, due to fiber ``reinforcement'' along the directions of oriented collagen fiber bundles.

Haughton and Merodio [2009] and Merodio and Haughton [2010] recently proposed a strain-energy density leading to localized strain softening, in order to
model the unusual mechanical behavior of tissues with Marfan syndrome. 
The influence of localized strain softening on the bifurcation of inflated cylinders under axial loading was studied and related to aneurysm formation, with particular reference to the mechanical response of arteries weakened by Marfan syndrome. 
It was shown that LSS  materials trigger bulging solutions. 
Haughton and Merodio's strain-energy density reads
\be \label{marfan}
W = \dfrac{\mu}{2} \left[\lambda_1^2 + \lambda_2^2 + \lambda_3^2 - \ee^{- \nu(\lambda_1^2 + \lambda_2^2 + \lambda_3^2 -3)^2} - 2 \right],
\en
where $\mu$ is the initial shear modulus, $\nu \ge 0$ is the \emph{localized strain softening (LSS) parameter}, and $\lambda_1$, $\lambda_2$, $\lambda_3$ are the principal stretch ratios (the square roots of the eigenvalues of the Cauchy-Green strain tensors).
Note that this strain energy density reduces to the neo-Hookean strain energy, $W = \mu(\lambda_1^2 + \lambda_2^2 + \lambda_3^2-3)/2$, when $\nu=0$.
\color{black}
We also remark that some recent experimental stress-strain tensile curves obtained for mouse models of Marfan syndrome do not seem to exhibit LSS [Eberth \emph{et al.}, 2009].
In effect, the precise mechanical properties of tissues affected by Marfan's syndrome are still a current topic of intense investigation.
Here we simply focus on the stability analysis of materials which may be modeled by the strain energy \eqref{marfan}.
\color{black}

\begin{figure}
  \centering 
  \subfigure{\epsfig{figure=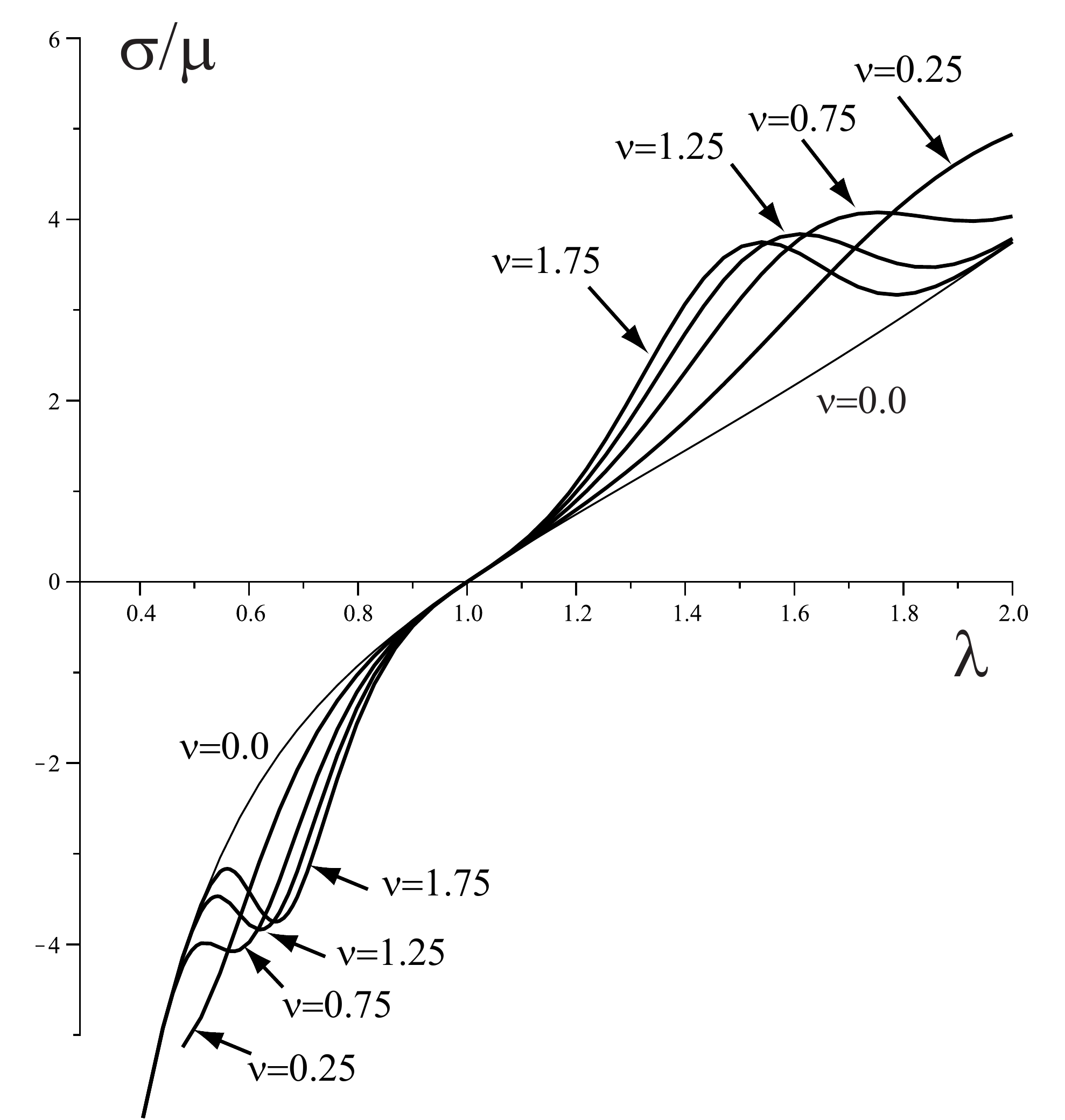, width=.48\textwidth}}
  \quad 
    \subfigure{\epsfig{figure=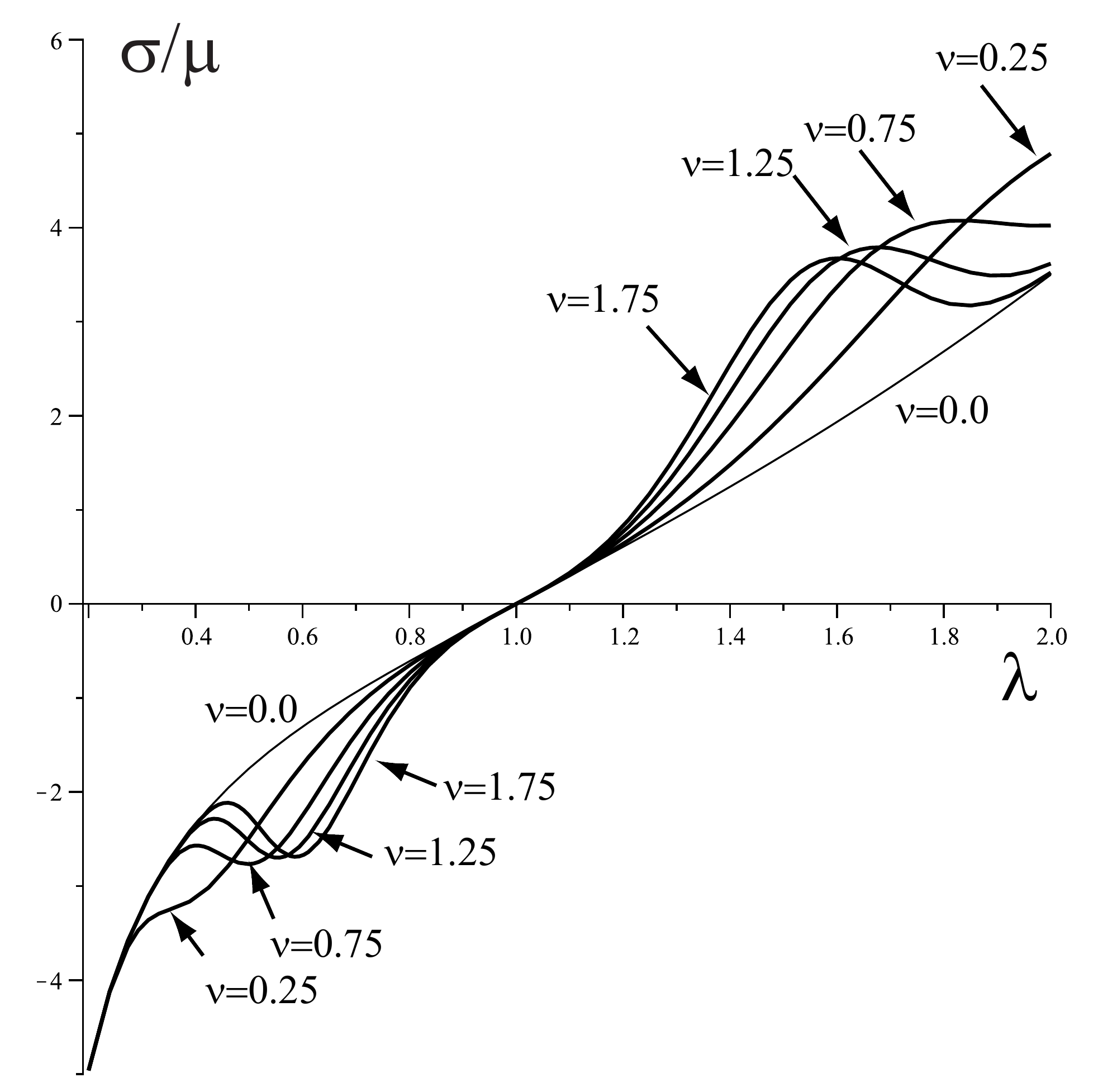, width=.48\textwidth}}
 \caption{Effect of localized strain softening for the large deformations of solids with strain-energy density given by \eqref{marfan}. 
 Stress-stretch curves in compression ($\lambda<1$) and in tension ($\lambda>1$), for plane strain deformation (figure on the left) and uni-axial deformation (figure on the right), with $\nu=0$ (neo-Hookean solid), and $\nu=0.25, 0.75, 1.25, 1.75$.}
 \label{fig:deformations}
\end{figure}

To illustrate how LSS affects the mechanical behavior of a solid with strain-energy \eqref{marfan}, compared to that of a neo-Hookean solid, we consider some common types of \emph{large static deformations}.
We call ($x_1$, $x_2$, $x_3$) the rectangular coordinate system aligned with the principal axes of the Cauchy-Green strain ellipsoid, and we examine in turn how the solid reacts to a \emph{plane strain deformation}  in the $x_1$-direction, with principal stretches
\be \label{plane}
\lambda_1 = \lambda, \qquad 
\lambda_2 = \lambda^{-1}, \qquad
\lambda_3=1,
\en
say, then to an \emph{equi-biaxial deformation} in the $x_1$- and $x_3$-directions, with principal stretches
\be \label{bi}
\lambda_1 = \lambda, \qquad 
\lambda_2 = \lambda^{-2}, \qquad
\lambda_3= \lambda,
\en
and finally, to a \emph{uni-axial deformation} in the $x_1$-direction, with principal stretches
\be \label{uni}
\lambda_1 = \lambda, \qquad 
\lambda_2 = \lambda^{-1/2}, \qquad
\lambda_3= \lambda^{-1/2}.
\en
Figure \ref{fig:deformations} shows the stress response in the cases \eqref{plane} and \eqref{uni}, with plots of the principal Cauchy stress $\sigma = \sigma_1$ as a function of the stretch $\lambda = \lambda_1$. 
Clearly, the LSS parameter is an influential factor of loss of convexity for these deformations, both in compression and in tension, as it leads to the apparition of small windows of strain-softening.


\color{black}
Our objective here is to 
\color{black}
study the stability of solids with strain energy \eqref{marfan} when they are subject to large compressions, by considering the possibility of superposed small-amplitude (i.e. \emph{incremental}) static solutions. 
This is the so-called \emph{Euler criterion} of stability [Beatty, 1996].
We restrict ourselves to \emph{two-dimensional} incremental deformations, for which the associated mechanical displacement $\vec{u}$ depends on two spatial coordinates only.
With respect to stability analysis, three quantities $\alpha$, $\beta$, $\gamma$ play an important role.
They are given in general by [Dowaikh and Ogden, 1990] 
\begin{align}
& \alpha = \lambda_1^2 \left(\lambda_1 W_1 - \lambda_2 W_2\right)/(\lambda_1^2 - \lambda_2^2) ,
\qquad
\gamma = \alpha \lambda_1^{-2}\lambda_2^2,
\\ \notag
&
2\beta = \left(W_{11} \lambda_1^2 + \lambda_2^2 W_{22}\right)/2 - \lambda_1 \lambda_2 W_{12} + \lambda_1 W_1 - \alpha,
\end{align}
where $W_i = \partial W/\partial \lambda_i$, $W_{ij} = \partial^2 W/\partial \lambda_i \lambda_j$ ($i,j=1,2$).
\color{black}
These quantities have the dimensions of elastic stiffnesses and play a role similar to that of elastic constants in an anisotropic material with orthorhombic symmetry. 
The difference here is that the anisotropy is strain-induced, not intrinsic and fixed, so that the values of $\alpha$, $\beta$, $\gamma$ change as the stretches $\lambda_1$, $\lambda_2$, $\lambda_3$ change under increasing compression.
\color{black}
For the strain energy \eqref{marfan}, we find that they are given by 
\begin{align} \label{alpha}
& \alpha = \mu \lambda_1^2 \left[ 1 + 2\nu(I_1-3)\ee^{-\nu(I_1-3)^2}\right],
\\ \notag
& \gamma =  \mu \lambda_2^2 \left[ 1 + 2\nu(I_1-3)\ee^{-\nu(I_1-3)^2}\right],\\ \notag
&
2\beta = \alpha + \gamma +  4\mu \nu(\lambda_1^2 - \lambda_2^2)^2 \left[ 1 - 2\nu(I_1-3)^2 \right]\ee^{-\nu(I_1-3)^2},
\end{align}
with $I_1 = \lambda_1^2 + \lambda_2^2 + \lambda_3^2$.

In Section \ref{Material-stability}, we study the {\color{black}bulk \color{black}  stability} of an infinite solid with strain-energy \eqref{marfan} \color{black}under compression.\color{black}
We find that, in contrast to the neo-Hookean solid, such a solid can become internally unstable at sufficiently high compression ratios, at least as long as $\nu$ is away from 0.0 (roughly, for $\nu > 0.5$).
Then, in Section \ref{Surface-stability}, we turn to the simplest \color{black} boundary value problem, with the stability analysis \color{black} of a semi-infinite solid under compression. 
As expected, we find that surface instability occurs earlier than \color{black}bulk \color{black}   stability, at critical compression ratios which are above those of a neo-Hookean solid (again, when $\nu$ is away from 0, say $\nu>0.1$).
These two sections suggest that solids with strain energy \eqref{marfan} are much more unstable in compression than classical hyperelastic solids. 
However, the picture is more nuanced when we turn our attention to problems of compressive instability in solids with \emph{finite size}.

For the compression of rectangular slabs (Section \ref{Slab-stability}) or of cylindrical circular tubes (Section \ref{Tube-stability}), we find hardly any difference with the neo-Hookean case, at least for slender bodies with dimensions compatible with the geometry of biological tissues. 
Only for the bending instability problem (Section \ref{Bending-stability}) do we find a significant difference with the neo-Hookean case. 
This can be explained by the known close relation between bending and surface stability [Gent and Cho, 1999], thus making the investigation of Section  \ref{Surface-stability} worthwhile a posteriori.
\color{black}For those finite size stability prototypes, we use dimensions which are representative of human arterial tissue dimensions.

We focus on compression stability issues because arteries are often compressed, for instance when they contract or when they are bent at tortuous sites.
In fact, the compression of the right coronary artery by a thoracic aneurysm of the ascending aorta has been reported for a patient with Marfan syndrome [Minami \emph{et al.}, 2007].
The separate problem of stability under large (lumen) pressure is treated elsewhere, see for example Merodio and Haughton [2010] or Han [2007].
 \color{black}  

The tentative conclusion of this paper is that for an instability occurring at low levels of compression in a neo-Hookean solid (such as for the compression of slender bodies, Sections \ref{Slab-stability} and \ref{Tube-stability}), the LSS effect does not play an influential role.
In contrast, it makes a solid with strain-energy density \eqref{marfan} more unstable than neo-Hookean solids for instabilities occurring at high levels of compression (Sections \ref{Material-stability},  \ref{Surface-stability}, \ref{Bending-stability}).


\section{\color{black}Bulk \color{black}   stability}
\label{Material-stability}


To investigate \emph{\color{black}bulk \color{black}   stability} (also called \emph{internal stability} by Biot [1965]), we study the ellipticity, or lack thereof, of the differential equations governing the equilibrium of small-amplitude two-dimensional static solutions, existing in the neighborhood of a large compressive strain \color{black} of a solid of infinite extend. \color{black}
Dowaikh and Ogden [1990] show that those equations are strongly elliptic when 
\be \label{material}
\alpha > 0, \qquad \beta + \sqrt{\alpha \gamma} > 0.
\en

Clearly, in view of the expression \eqref{alpha}$_1$ for $\alpha$, the first inequality is always satisfied for solids with strain energy \eqref{marfan}, as is the case for neo-Hookean solids. 
However, in contrast to those latter solids, the other inequality may or may not be satisfied, depending on the magnitude of the LSS parameter $\nu$.
To illustrate this possibility, we specialize the analysis to certain common large deformations.

\begin{figure}
\center
\epsfig{figure=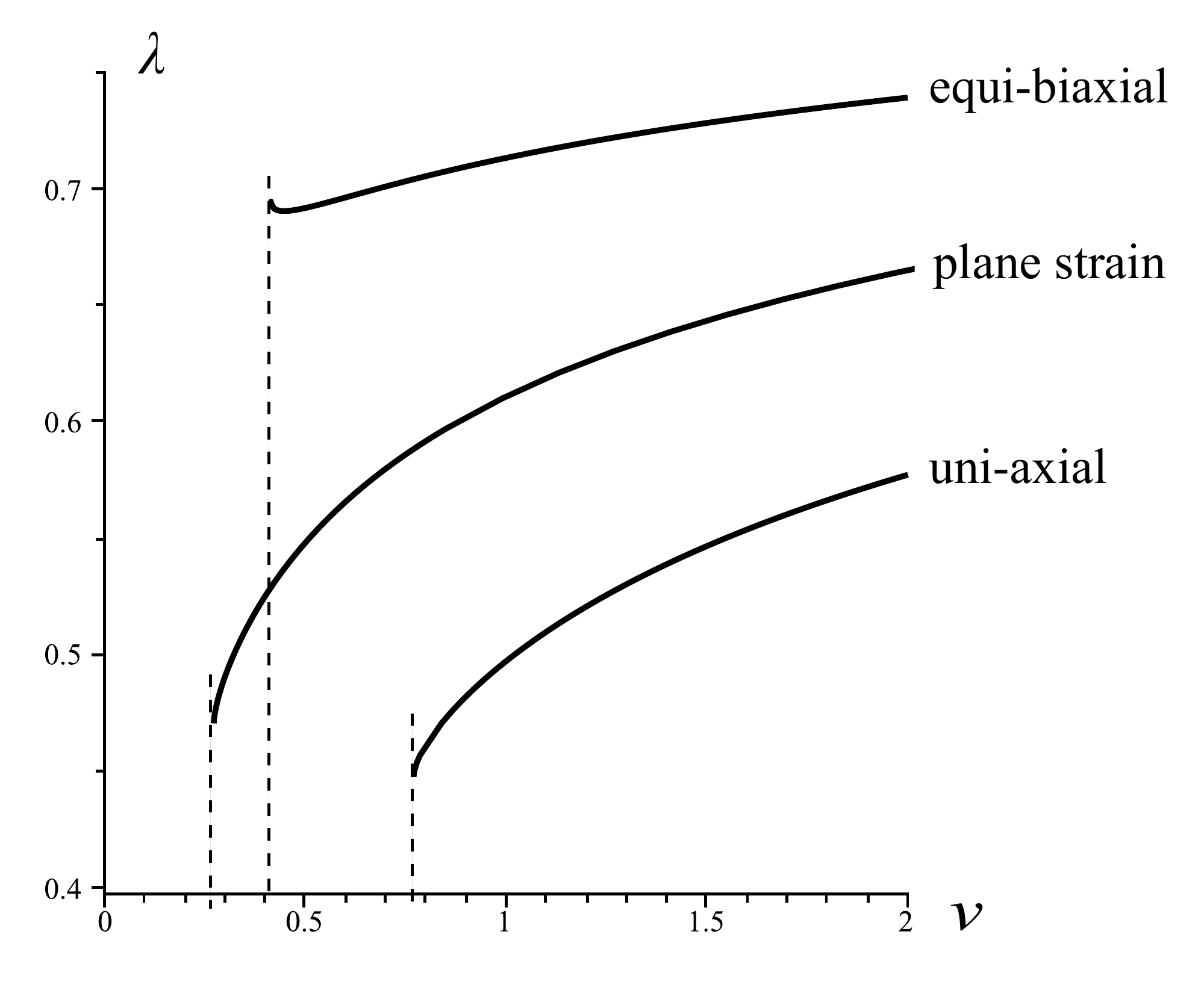, width=.66\textwidth}
 \caption{Loss of ellipticity (internal instability) for a compressed infinite solid with strain-energy density given by \eqref{marfan}. 
 For a given stretch ratio  $\lambda$ of uni-axial compression,  plane strain compression, or equi-biaxial compression, and for a solid characterized by a given $\nu$, the incremental equations of elastostatics loose their ellipticity when a point with coordinates ($\lambda$, $\nu$) is situated below the thick curves.  
 For the neo-Hookean solid ($\nu=0$), there is no loss of ellipticity.}
 \label{fig:bulk}
\end{figure}

First, take the large strain to be a \emph{plane strain compression} \eqref{plane} in the $x_1$-direction.
Then, a numerical study of the variations of the quantity $\beta + \sqrt{\alpha\gamma}$ reveals that there exists a certain value $\nu_m$ such that for $\nu \ge \nu_m$, there always exist values of $\lambda$ where this quantity is zero.
This delimits a curve in the ($\lambda,\nu$)-plane beyond which ellipticity is lost.
We find that this scenario is repeated for the \emph{equi-biaxial compression} \eqref{bi} in the $x_1$- and $x_3$-directions, 
and for the \emph{uni-axial compression} \eqref{uni} in the $x_1$-direction.
In particular, we find that $\nu_m \simeq 0.272$, $0.415$, and $0.770$, for the deformations \eqref{plane}, \eqref{bi}, and \eqref{uni}, respectively, see Figure \ref{fig:bulk}.


\section{Surface stability}
\label{Surface-stability}


Now take a semi-infinite solid with strain-energy density \eqref{marfan}.
In this section, we study surface instability, and thus consider that the solid occupies the $x_2 \ge 0$ region in its deformed state, with the boundary $x_2=0$ free of traction.

Biot [1963] shows that when a half-space is subjected to a large homogeneous deformation, there eventually exists a configuration where it is possible to superpose an infinity of small-amplitude static perturbations, with amplitude variations confined to the neighborhood of boundary. 
He considers that this configuration corresponds to the surface buckling, or surface instability, phenomenon. 
Dowaikh and Ogden [1990] find that, for a general isotropic incompressible solid, the \emph{surface stability criterion} can be put in the form
\be \label{surf}
\gamma > 0, \qquad 
\alpha - \gamma + 2 \sqrt{\alpha/\gamma} \ (\beta + \gamma)>0.
\en

Clearly, in view of the expression \eqref{alpha}$_2$ for $\gamma$, the first inequality is always satisfied for solids with strain energy \eqref{marfan}, as is the case for neo-Hookean solids. 
Moreover, we recall that [Dowaikh and Ogden, 1990], if the second inequality is satisfied here, then the second inequality in \eqref{material} is automatically satisfied. 
In other words, surface instability always precedes internal (\color{black}bulk\color{black}) instability.

\begin{figure}
\center
\epsfig{figure=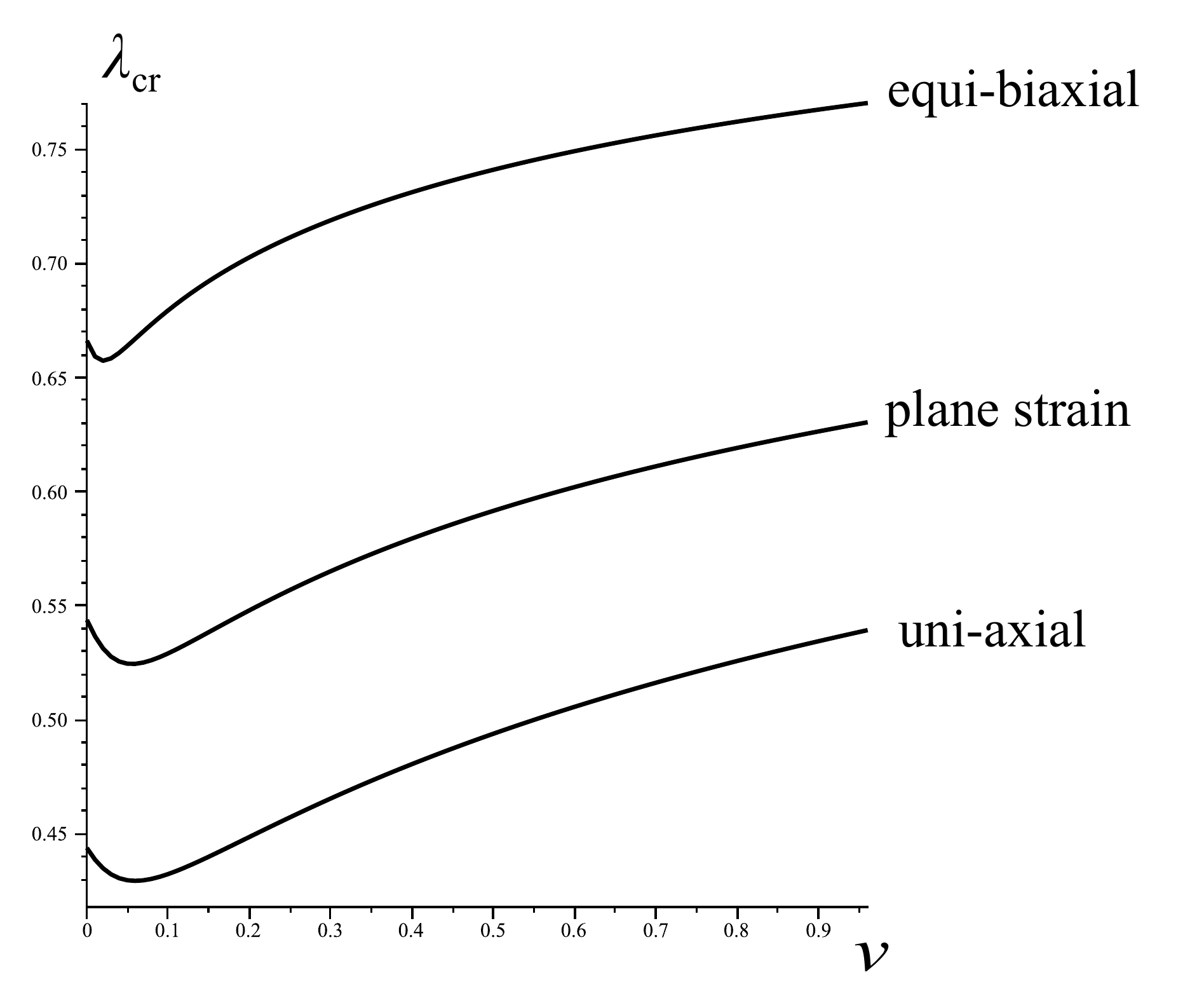, width=.7\textwidth}
 \caption{Surface instability of a compressed solid with strain-energy density given by \eqref{marfan}. 
 For uni-axial compression,  plane strain compression, and equi-biaxial compression, the surface instability occurs earlier than for the neo-Hookean solid (which corresponds to $\nu=0$), once $\nu > 0.15$.}
 \label{fig:surface}
\end{figure}

We use criterion \eqref{surf}$_2$ as an equality to compute the critical stretch of compression $\lambda_\text{cr}$ for surface instability, in the cases of \emph{plane strain}, as described by \eqref{plane}; of \emph{equi-biaxial compression}, as described by \eqref{bi}; and of \emph{uni-axial compression}, as described by \eqref{uni}.
Figure \ref{fig:surface} shows the resulting plots of $\lambda_\text{cr}$ as a function of the  LSS parameter $\nu$. 
At $\nu=0$, we recover the critical stretch ratios of the neo-Hookean solid [Biot, 1963], i.e. 0.44 in uni-axial compression, 0.54 in plane strain, and 0.66 in equi-biaxial compression.
Then, there is a narrow range of values for $\nu \gtrapprox 0$ where the half-space is slightly more stable than a neo-Hookean half-space.
However, when $\nu$ is greater than $0.15$, we find that the half-space becomes unstable at higher compression ratios (earlier) than in the neo-Hookean case.
This situation is in complete contrast with the isotropic models usually employed to model strain-stiffening biological soft tissues, which are always much more stable than the neo-Hookean solid with respect to surface buckling (see Goriely \emph{et al.} [2006] and references therein).

These results, and those of the previous section, suggest that tissues modeled with strain-energy density \eqref{marfan} become easily unstable when compressed. 
At the same time, it must be kept in mind that an infinite or semi-infinite medium does not exist and is an idealization; in Nature, all solids have finite dimensions.
In the next sections, we investigate the influence of finite size for some prototype stability problems.


\section{Slab stability}
\label{Slab-stability}


Here we consider the homogeneous compression of a thick slab until it buckles (incrementally).
Among others, Biot  [1963b], Ogden and Roxburgh [1993], and Beatty and Pan [1998] have determined the bifurcation criterion in the case of a material with neo-Hookean strain-energy density. 
It reads
\be
\dfrac{\tanh \left(\lambda_1\lambda_2^{-1} k h\right)}{\tanh(kh)} = \left[\dfrac{1+\lambda_1^2\lambda_2^{-2}}{4\lambda_1\lambda_2^{-1}}\right]^{\pm 1},
\en
when $\lambda_1$ is the principal stretch in the direction of compression.
Here the plus (minus) sign is associated with antisymmetric/flexural (symmetric/extensional) modes of buckling, and $k h$ gives a measure of the current slenderness of the slab. 
Explicitly, $k h = n \pi h/l$, where $n$ is an integer, $2h$ is the current width of the slab (in the $x_1$ direction), and $2l$ is its current height (in the $x_2$ direction).
In terms of the slab's original width $2H$ and length $2L$ (say), we have
\be \label{kh}
k h = n \pi (\lambda_2/\lambda_1)(H/L).
\en

Ogden and Roxburgh [1993] also give explicit bifurcation criteria for a general hyperelastic material, but there are several of these depending on whether the buckling is antisymmetric or symmetric and on the nature (purely imaginary or complex) of the roots to the following biquadratic:
\be \label{biquadratic}
\gamma q^4 + 2 \beta q^2 + \alpha = 0.
\en
For neo-Hookean solids, these roots are $q_{1,2} = \pm \ii$, $q_{3,4} = \pm \ii \lambda_1 \lambda_2^{-1}$  and their nature is fixed. 
For solids with strain-energy density  \eqref{marfan} however, the nature of the roots is strongly dependent of the value of the physical parameter $\nu$ and of the geometrical parameters $\lambda_1$, $\lambda_2$. 
Instead of discussing the different possibilities, we turn to the Stroh formalism to produce a numerically stable bifurcation criterion, independent of the nature of the roots and of the buckling mode. 

First, we recall that the incremental equations of equilibrium can be put in the following form:
\be \label{stroh}
\vec{\eta}(x_2)' = \ii k N \vec{\eta}(x_2),
\en
where $\vec{\eta}$ is the displacement-traction vector and $N$ the Stroh matrix. 
Explicitly, taking the mechanical displacement $\vec{u}$ and the nominal traction $\vec{s}$ on the faces $x_2=\text{const.}$ in the form
\be
\{ \vec{u}, \vec{s} \} = \left\{ \vec{U}(x_2), \ii k \vec{S}(x_2) \right\} \ee^{\ii k x_1},
\en
where $\vec{U}$ and $\vec{S}$ are single-valued functions, we define $\vec{\eta}$ as $\vec{\eta} = [\vec{U}, \vec{S}]^t$.
On the other hand, $N$ is given, in the case where the $x_2=\pm h$ faces of the slab are free of traction, by [Destrade \emph{et al.}, 2005]
\be
N =
\begin{bmatrix}
0 & -1 & 1/\gamma & 0 \\
-1 & 0 & 0 & 0 \\
-(2\beta + \gamma) & 0 & 0 & -1 \\
0 & \alpha - \gamma & -1 & 0
\end{bmatrix}.
\en

Because $N$ is a constant matrix, the general solution to \eqref{stroh} is of the form $\vec{\eta}(x_2) = \mathcal{N}(x_2)\vec{b}$, where $\vec{b}$ is a constant vector and $\mathcal{N}$ is a fundamental matrix solution:
\be
\mathcal{N}(x_2) = \left[ \vec{\zeta}^1 |\vec{\zeta}^2 |\vec{\zeta}^3 |\vec{\zeta}^4 \right]\text{Diag}\left(\ee^{\ii k q_1 x_2}, \ee^{\ii k q_2 x_2}, \ee^{\ii k q_3 x_2}, \ee^{\ii k q_4 x_2} \right),
\en
with the $q$'s and $\vec{\zeta}$'s eigenvalues and eigenvectors, respectively, of the eigenproblem $N \vec{\zeta} = q \vec{\zeta}$.
Explicitly, the $q$'s are the four roots of the characteristic equation $\det (N - qI)=0$, which is the quartic \eqref{biquadratic}, and the $\vec{\zeta}$'s are [Destrade, 2007] 
\be
\vec{\zeta}^j = \left[ - q_j^2/\gamma, q_j/\gamma, - q_j(q_j^2-1), q_j^2 - \alpha/\gamma \right]^t, \qquad j = 1, \ldots, 4.
\en

We may now write the solution on one face of the slab (at $x_2=h$) in terms of the solution on the other face (at $x_2=-h$) as: $\vec{\eta}(h) = M \vec{\eta}(-h)$, with $M$, the matricant [Shuvalov, 2000], defined as
\be
M = \mathcal{N}(h)\left[\mathcal{N}(-h)\right]^{-1} = \begin{bmatrix} M_1 & M_2 \\ M_3 & M_4 \end{bmatrix},
\en
where $M_1$, etc. are $2 \times 2$ submatrices. 
The incremental boundary conditions are that the end faces of the slab stay in sliding contact with the thrust platens: this is ensured by \eqref{kh}; and that its lateral faces remain free of traction: $\vec{S}(h) = \vec{S}(-h) = \vec{0}$.
These identities lead to the robust form of the bifurcation criterion, 
\be \label{M3}
\ii \dfrac{\det  M_3}{\det  M_1} = 0,
\en 
and the left hand side of this equation is always real [Shuvalov, 2000] prior to buckling, independently of the nature of the roots $q$.

\begin{figure}
\center
\epsfig{figure=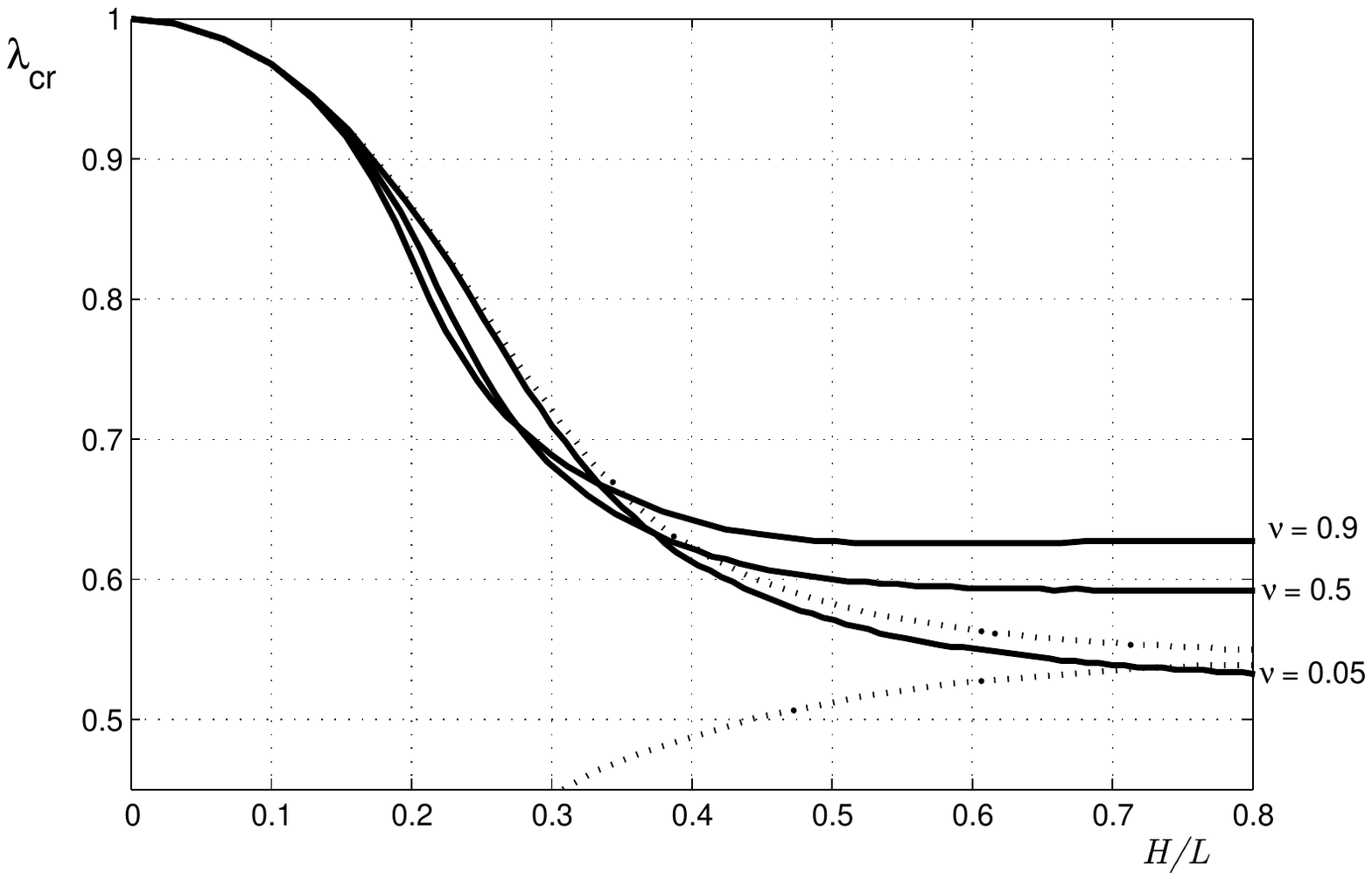, width=.7\textwidth}
 \caption{Buckling of a compressed slab in plane strain: critical stretch against $H/L$, the slab's initial slenderness.
 Solid curves: solids with strain-energy density given by \eqref{marfan}. 
 Dashed curves: neo-Hookean solids.}
 \label{fig:slab}
\end{figure}

We implemented this criterion for the case treated by Biot [1963] that is, plane strain compression \eqref{plane}, when the slab is not allowed to expand in the $x_3$-direction. 
In Figure \ref{fig:slab}, we compare the critical stretches of compression for slabs made of solids with strain-energy density \eqref{marfan} (solid plots) with those for neo-Hookean slabs (dashed plots).
In the latter case we obtain two curves, the upper one corresponding to antisymmetric (flexural) modes of buckling and the lower one to  a symmetric (compressional) mode, which is never attained, because it lies below the plots for antisymmetric buckling modes.
In the former case, there is a single curve (solid line) for each value of the LSS parameter $\nu$, corresponding to the single bifurcation criterion \eqref{M3}.

We plotted the value of the critical stretch of compression $\lambda_\text{cr}$ against the initial slenderness of the slab $H/L$, for the range $0 \le H/L \le 0.8$, which is likely to cover all realistic dimensions of living tissues.
We show the plots obtained at $n=1$, but not those for $n \ge 2$, because they are all below the $n=1$ antisymmetric mode, and are thus never reached.
We see that in the infinitely thick slab limit $H/L \rightarrow \infty$, the curves tend to the asymptotic value given by the critical stretch of surface stability of the previous section. 
In that limit, solid slabs with strain-energy given by \eqref{marfan} buckle earlier than neo-Hookean slabs when $\nu > 0.15$, in agreement with the results of Figure \ref{fig:surface}.
However soft tissues are in general quite slender and we note that in the low $H/L$ region ($0 \le H/L \le 0.2$), the differences between the different curves become minute.
In fact, the solids with strain-energy density \eqref{marfan} turn out to be more stable than neo-Hookean solids in the $0.2 \le H/L \le 0.3$ range.


\section{Tube stability}
\label{Tube-stability}


Now we study the stability of a compressed tube.
When it is placed between two lubricated platens, an elastic tube deforms homogeneously (in an equi-biaxial manner), and an exact incremental solution exists, due to Wilkes [1955].

The resolution of the stability problem can be conducted along lines similar to those outlined in the previous section, as shown by Goriely \emph{et al.} [2008].
The main differences are that now the displacement and traction are sought in the form:
\be
\{ \vec{u}, \vec{s} \} = \left\{ \vec{U}(r), \ii k \vec{S}(r) \right\} \ee^{\ii k \theta},
\en
where $r$, $\theta$ are the cylindrical coordinates in the deformed configuration; the displacement-traction vector is defined as $\vec{\eta}  = [\vec{U}(r), r\vec{S}(r)]^t$ and it has six components instead of four; and the cylindrical counterpart to the rectangular incremental equations of equilibrium \eqref{stroh} are
\be \label{stroh-tube}
\dfrac{\text{d}\vec{\eta}}{\text{d}r}(r) = \dfrac{\ii}{r} N(r) \vec{\eta}(r),
\en
where the $6 \times 6$ Stroh matrix now has variable components (see Goriely \emph{et al.} [2008] for their explicit expression). 
The eigenvectors of $N$ are written in terms of Bessel functions. 
The resulting bifurcation criterion is again of the form \eqref{M3}.

\begin{figure}
\center
\epsfig{figure=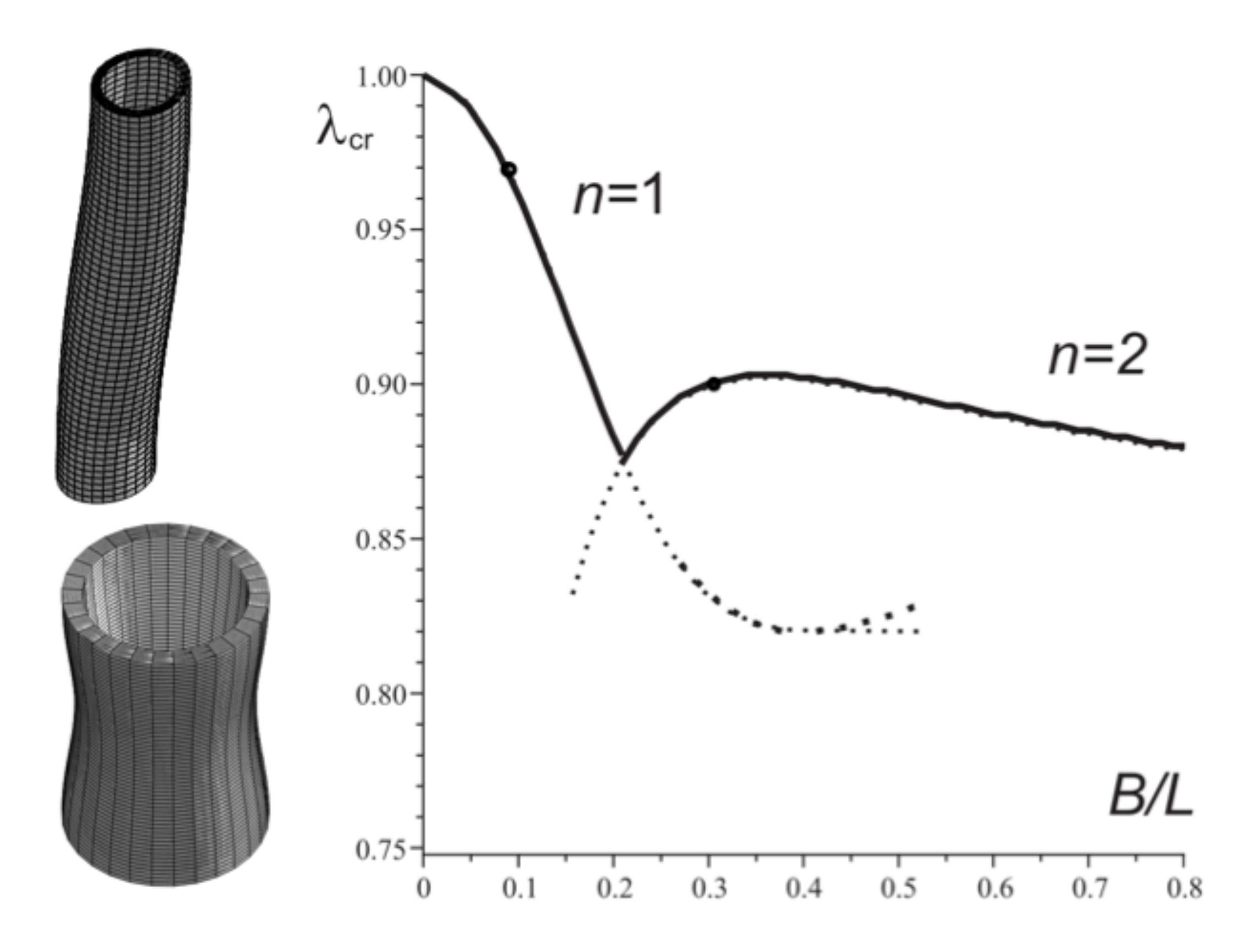, width=.7\textwidth, height=.55\textwidth}
 \caption{Buckling of a compressed tube with dimensions close to those of a human carotid artery: $B/A=1.25$ and $0.0 \le B/L \le 0.8$, where $A$, $B$, $L$ are the initial inner radius, outer radius, and length of the tube. 
Here we traced the bifurcation plots for $\nu = 0.0$ (neo-Hookean), 0.25, 0.5, 0.9, and found that all four superposed almost exactly.
\color{black}
The figures on the left show the buckling modes corresponding to tubes with slenderness 0.1 (top figure) and 0.3 (bottom figure), denoted by two black spots on the dispersion curve. \color{black}}
 \label{fig:tube}
\end{figure}

The main tubular structures in the body are arteries, and we focus on this class of geometrical objects. 
We find in Delfino \emph{et al.} [1997] that in their unloaded configuration, human carotid arteries have a wall thickness ranging from 0.4 to 0.9 mm, and inner diameter ranging from 1.8 to 3.3 mm.
For our calculations we thus take $B/A = 1.25$  as being a representative geometric parameter, where $A$ and $B$ are the initial inner and outer radii, respectively.
We then implement the numerical procedure of Goriely \emph{et al.} [2008] to determine the critical stretch of compression for tubes with slenderness $B/L$ in the $0.0-0.8$ range, where $L$ is the initial tube length.
We find that in that range, the bifurcation curves corresponding to solids with LSS parameter $\nu = 0.25$, 0.5, 0.9, all superpose almost exactly with the bifurcation curve of the neo-Hookean tube (at $\nu = 0.0$) and are indistinguishable one from another.
For slender tubes, $0.0 \le B/L \le 0.21$, the buckling occurs in the antisymmetric $n=1$ mode.
For stubbier tubes, $0.21 \le B/L \le 0.8$, the barreling mode $n=2$ is preferred.
The curves of all other modes ($n=0$ and $n \ge 3$) are all situated below these two.
Practically, it means that the tube cannot be compressed axially by more than 13\% (because $\lambda_\text{cr}=0.87$ at $B/L = 0.21$), at least when its slenderness $B/L$ is less than 0.8.
At that low level of strain, the solid with strain energy \eqref{marfan} behaves almost exactly like a neo-Hookean solid, as can be checked by inspection of the curves in Figure \ref{fig:deformations}. 
This explains the coincidence of the curves in Figure \ref{fig:tube}. 
Only in the low part of the $n=1$ curve can a distinction be made, but it is never expressed because the $n=2$ buckling mode is prevalent at the corresponding ratios.


\section{Bending stability}
\label{Bending-stability}


As a final example of compression instability, we study the effect of LSS on the bending stability criterion.
For neo-Hookean solids, this instability occurs at compression ratios very close to the critical compression ratio of surface instability in plane strain. 
Quantitatively, the critical value for the circumferential stretch on the inner face of a bent neo-Hookean block is $\lambda_\text{cr} \simeq 0.56$, which corresponds to only 2\% less compression than the surface stability critical ratio of 0.54, see Section 3.
At these levels of strain, the behavior of solids with strain-energy \eqref{marfan} is likely to diverge from that of neo-Hookean solids.

Take a rectangular block, made of a hyperelastic solid initially contained in the region 
\be
0 \le X_1 \le L, \qquad
-A \le X_2 \le A, \qquad
0 \le X_3 \le H,
\en
say. 
Then bend it into the cylindrical sector located in the region
\be
r_a \le r \le r_b, \qquad
-\alpha \le \theta \le \alpha, \qquad
0 \le z \le H,
\en
where ($r, \theta, z$) are the cylindrical coordinates in the current configuration; $\alpha$, the angle of bending, is prescribed; and the inner and outer radii, $r_a$ and $r_b$, respectively, can be found from Rivlin's bending deformation [Rivlin, 1949]:
\begin{equation} \label{RivlinSln}
r = \sqrt{2 L X/\alpha +  (r_a^2 + r_b^2)/2}, \qquad 
\theta = \alpha Y / L, \qquad z = Z.
\end{equation}
Destrade \emph{et al.} [2010] show that
\be
r_a^2, r_b^2 = \dfrac{L}{\alpha}\left(\sqrt{4A^2+\dfrac{L^2}{\alpha^2}} \mp 2A\right).
\en
Calling $\lambda_2$ the circumferential stretch, Destrade \emph{et al.} [2010] also establish that it varies through the wall thickness from value $\lambda_2^a$ at $r=r_a$ to value $\lambda_2^b$ at $r=r_b$, given by 
\begin{equation} \label{lambda_ab}
\lambda_2^a = \sqrt{\sqrt{1+ (2 A\alpha/L)^2} -  (2 A\alpha/L)}, \qquad
\lambda_2^b = \sqrt{\sqrt{1+ (2 A\alpha/L)^2} + (2 A\alpha/L)} = 1/\lambda_2^a. 
\end{equation}

Now consider the possibility of a superposed incremental solution, describing the apparition of prismatic wrinkles on the inner face of the bent block. 
Destrade \emph{et al.} [2009] show that the corresponding incremental equations of equilibrium can be put in the following fully non-dimensional Stroh form:
\be \la{stroh3}
\dfrac{\text{d}}{\text{d} \lambda_2} \vec{\eta}(\lambda_2) = \dfrac{\ii}{\lambda_2} \vec{N}(\lambda_2) \vec{\eta}(\lambda_2),
\en
where the incremental displacement-traction vector $\vec{\eta}$ and the Stroh matrix $N$ are seen as functions of $\lambda_2$ only. 
Explicitly,
\be 
\vec{N} = \begin{bmatrix} 
  \ii & -n & 0 & 0 \\
  -n (1 - \sigma/\alpha ) &  - \ii (1 - \sigma/\alpha) & 0  &  -  1/\alpha\\
  \kappa_{11} & \ii \kappa_{12} & -\ii & -n (1 - \sigma/\alpha) \\
  - \ii \kappa_{12} & \kappa_{22}& -n &  \ii (1 - \sigma/\alpha)  
          \end{bmatrix},
\en
where
\begin{align}
& \kappa_{11} = 2(\beta + \alpha - \sigma) 
  + n^2\left[\gamma - (\alpha - \sigma)^2/\alpha\right],
 \notag \\[2pt]
& \kappa_{12} = n \left(2 \beta + \alpha + \gamma - \sigma^2/\alpha\right),
 \notag \\[2pt]
& \kappa_{22} =  
   \gamma - (\alpha - \sigma)^2/\alpha
   + 2 n^2(\beta + \alpha - \sigma),
\end{align}
and the quantities $\gamma$, $\alpha$, and $\beta$ are given in \eqref{alpha}, by taking the plane strain configuration \eqref{plane}.
Also, $\sigma$ is the radial component of the Cauchy stress  necessary to maintain the block in the bent configuration, with inner and outer faces free of traction:
\be
\sigma = W(1/\lambda_2, \lambda_2, 1) - W(1/\lambda_2^a, \lambda_2^a, 1),
\en
and $n$ is the circumferential number, determined from the condition that there are no incremental normal tractions on the end faces $\theta = \pm \alpha$: it is such that [Haughton, 1999]
\be
n = p \pi / \alpha,
\en
for some integer $p$, the \emph{mode number}.

The differential system \eqref{stroh3} must be integrated from $\lambda_2^a$ to $\lambda_2^b$, subject to the incremental boundary conditions of traction-free bent faces:
 \be \la{BC}
 \vec{\eta} (\lambda_2^a) = [\vec{U}_a, \vec{0}]^t,
\qquad 
 \vec{\eta} (\lambda_2^b) = [\vec{U}_b, \vec{0}]^t,
\en
where $\vec{U}_a$, $\vec{U}_b$ are non-zero constant vectors.
The numerical treatment of this two-point boundary value problem is delicate due to potential stiffness issues. 
However, the compound matrix method (see [Haughton, 1999; Destrade \emph{et al.} 2009]) smoothens these out.

\begin{figure}
\center
\epsfig{figure=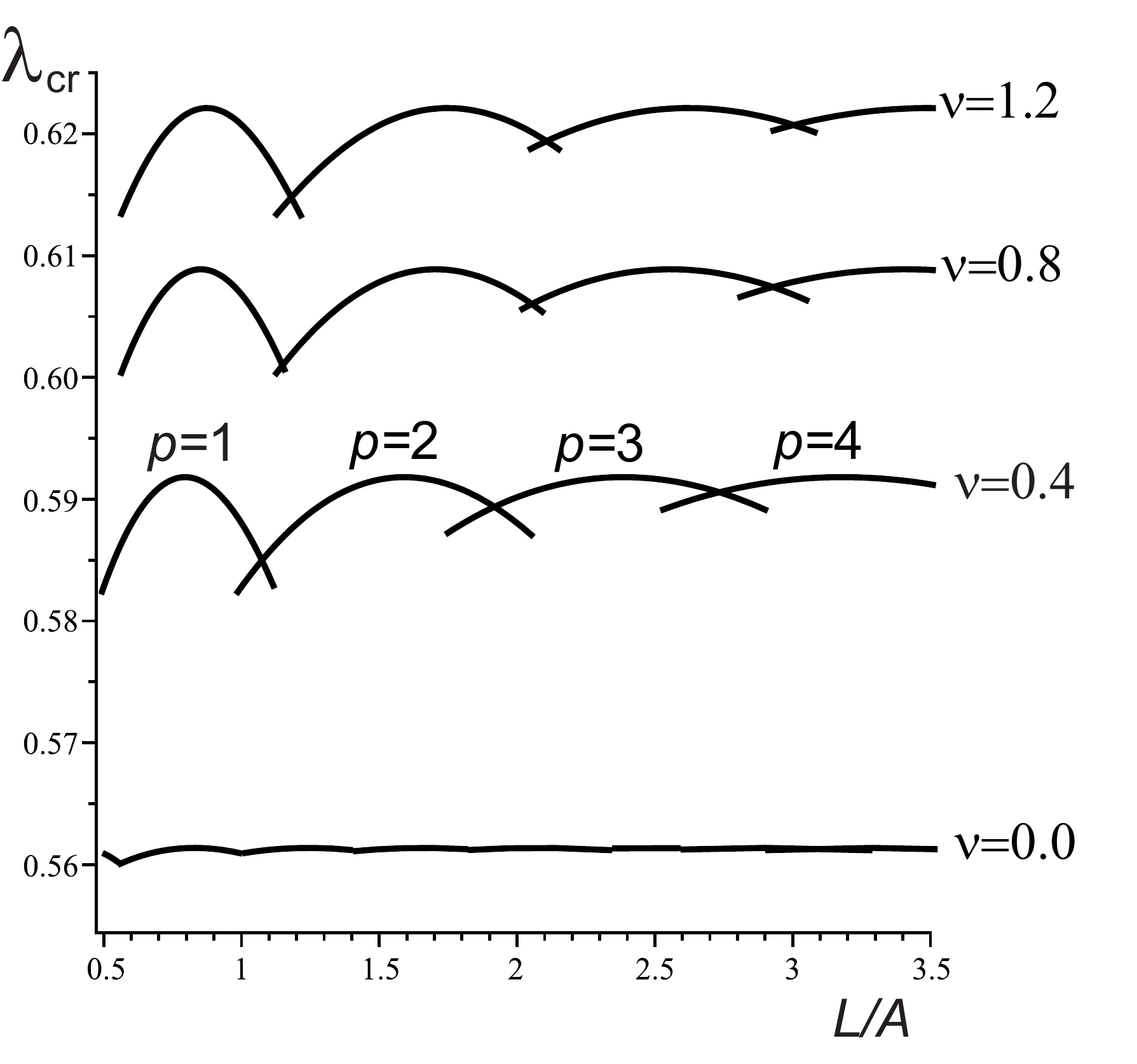, width=.7\textwidth, height=.55\textwidth}
 \caption{Bending instability for a rectangular block with length-to-depth ratio $L/(2A) \in [0.25, 1.75]$. 
The neo-Hookean solid ($\nu = 0.0$) can be bent more than solids with strain energy density \eqref{marfan}, see plots at $\nu = 0.4, 0.8, 1.2$.}
 \label{fig:bending}
\end{figure}

We implemented this technique to find the critical circumferential stretch ratio in terms of the block's aspect ratio $L/A$, see Figure \ref{fig:bending}.
At $\nu=0.0$, the neo-Hookean solid buckles at $\lambda_\text{cr} \simeq 0.56$, as previously shown [Haughton, 1999], with a mode number which increases rapidly as the block is more slender ($L/A$ increases).
For $\nu = 0.4$, we find that blocks with strain energy \eqref{marfan} buckle earlier, at $\lambda_\text{cr} \simeq 0.59$.
Similarly, $\lambda_\text{cr}$ increases as $\nu$ increases. 
Hence $\lambda_\text{cr} \simeq 0.61$ at $\nu=0.8$ and $\lambda_\text{cr} \simeq 0.62$ at $\nu=1.2$.
Also, we see that there are less wrinkles on the inner bent face than in the neo-Hookean case.
For example, a block which is 1.5 times longer than it is deep ($L/(2A) = 1.5$, giving $L/A=3$) buckles with $p=4$ wrinkles when $\nu>0.4$, in contrast with $p=7$ wrinkles in the neo-Hookean case [Destrade \emph{et al.}, 2009].

In conclusion, the level of compression at which instability occurs for the neo-Hookean solid is a good indicator of instability levels for solids with strain energy \eqref{marfan}.
If it occurs at high compression levels (as in this section), then the latter solids are unstable much earlier than neo-Hookean solids. 
Otherwise, (as in Sections \ref{Slab-stability} and \ref{Tube-stability}), there are no noticeable differences between the two types of solids.


\section*{Acknowledgments}

The first author gratefully acknowledges the support of a Senior Marie Curie Fellowship from the European Commission and the hospitality of Department of Continuum Mechanics and Structures at the 
E.T.S. Ing. Caminos, Canales y Puertos, Universidad Polit\'ecnica de Madrid.



\end{document}